\documentclass[twoside,onecolumn,9pt]{article}
\usepackage{extsizes}
\usepackage[super,sort&compress,comma]{natbib} 
\usepackage[version=3]{mhchem}
\usepackage[left=1.5cm, right=1.5cm, top=1.785cm, bottom=2.0cm]{geometry}
\usepackage{balance}
\usepackage{mathptmx}
\usepackage{sectsty}
\usepackage{graphicx} 
\usepackage{lastpage}
\usepackage[format=plain,justification=justified,singlelinecheck=false,font={stretch=1.125,small,sf},labelfont=bf,labelsep=space]{caption}
\usepackage{float}
\usepackage{fancyhdr}
\usepackage{fnpos}
\usepackage[english]{babel}
\addto{\captionsenglish}{%
  
}
\usepackage{array}
\usepackage{droidsans}
\usepackage{charter}
\usepackage[T1]{fontenc}
\usepackage[usenames,dvipsnames]{xcolor}
\usepackage{setspace}
\usepackage[compact]{titlesec}
\usepackage{hyperref}

\usepackage[outdir=./]{epstopdf}

\definecolor{cream}{RGB}{222,217,201}

\begin{document}

\pagestyle{fancy}
\thispagestyle{plain}
\fancypagestyle{plain}{
\renewcommand{\headrulewidth}{0pt}
}

\makeFNbottom
\makeatletter
\renewcommand\LARGE{\@setfontsize\LARGE{15pt}{17}}
\renewcommand\Large{\@setfontsize\Large{12pt}{14}}
\renewcommand\large{\@setfontsize\large{10pt}{12}}
\renewcommand\footnotesize{\@setfontsize\footnotesize{7pt}{10}}
\makeatother

\renewcommand{\thefootnote}{\fnsymbol{footnote}}
\renewcommand\footnoterule{\vspace*{1pt}%
\color{cream}\hrule width 3.5in height 0.4pt \color{black}\vspace*{5pt}} 
\setcounter{secnumdepth}{5}

\makeatletter 
\renewcommand\@biblabel[1]{#1}            
\renewcommand\@makefntext[1]%
{\noindent\makebox[0pt][r]{\@thefnmark\,}#1}
\makeatother 
\renewcommand{\figurename}{\small{Fig.}~}
\sectionfont{\sffamily\Large}
\subsectionfont{\normalsize}
\subsubsectionfont{\bf}
\setstretch{1.125} 
\setlength{\skip\footins}{0.8cm}
\setlength{\footnotesep}{0.25cm}
\setlength{\jot}{10pt}
\titlespacing*{\section}{0pt}{4pt}{4pt}
\titlespacing*{\subsection}{0pt}{15pt}{1pt}

\fancyfoot{}
\fancyfoot[RO]{\footnotesize{\sffamily{1--\pageref{LastPage} ~\textbar  \hspace{2pt}\thepage}}}
\fancyfoot[LE]{\footnotesize{\sffamily{\thepage~\textbar\hspace{4.65cm} 1--\pageref{LastPage}}}}
\fancyhead{}
\renewcommand{\headrulewidth}{0pt} 
\renewcommand{\footrulewidth}{0pt}
\setlength{\arrayrulewidth}{1pt}
\setlength{\columnsep}{6.5mm}
\setlength\bibsep{1pt}

\makeatletter 
\newlength{\figrulesep} 
\setlength{\figrulesep}{0.5\textfloatsep} 

\newcommand{\topfigrule}{\vspace*{-1pt}%
\noindent{\color{cream}\rule[-\figrulesep]{\columnwidth}{1.5pt}} }

\newcommand{\botfigrule}{\vspace*{-2pt}%
\noindent{\color{cream}\rule[\figrulesep]{\columnwidth}{1.5pt}} }

\newcommand{\dblfigrule}{\vspace*{-1pt}%
\noindent{\color{cream}\rule[-\figrulesep]{\textwidth}{1.5pt}} }

\makeatother

\begin{@twocolumnfalse}

		\noindent\LARGE{\textbf{Mixed Formamidinium-Methylammonium Lead Iodide perovskite from first-principles: Hydrogen-bonding impact on the electronic properties$^\dag$}} \\

		\noindent\large{Maximiliano Senno,\textit{$^{a}$} and Silvia Tinte$^{\ast}$\textit{$^{a,b}$}} \\

		\noindent\large{$^{a}$~Instituto de F\a'{\i}sica del Litoral, CONICET - Universidad Nacional del Litoral, G\"uemes 3450 (3000) Santa Fe, Argentina. }
		\noindent\large{$^{b}$~Facultad de Ingenier\a'{\i}a Qu\a'{\i}mica, Universidad Nacional del Litoral, Santiago del Estero 2829, (3000) Santa Fe, Argentina. E-mail: silvia.tinte@santafe-conicet.gov.ar}
		
		\vspace{0.5cm}
		
		\noindent\normalsize{Hybrid perovskites with mixed organic cations such as methylammonium (CH$_3$NH$_3$, MA) and formamidinium (CH(NH$_2$)$_{2}$, FA) have attracted interest due to their improved stability and capability to tune their properties varying the composition. 
		Theoretical investigations in the whole compositional range for these mixed perovskites are scarce in part due to the limitations of modeling cationic orientation disorder. In this work, we report on the 
		local 
		variation of the structural and electronic properties in mixed A-site cation MA/FA lead iodide perovskites FA$_x$MA$_{1-x}$PbI$_3$ evaluated  
                from static first-principles calculations in certain structures where the orientations of organic cations result from examining the energy landscape of some compositions.
                The cation replacement at the A-site to form the solid solution causes an increase tilting of the inorganic PbI$_6$ octahedra: in the FA-rich compounds the replacement of FA by a smaller cation like MA is to compensate the reduced space filling offered by the smaller cation, whereas in the MA-rich compounds is to expand the space needed for the larger cation. In fact, the effect of octahedron tiltings exceeds that of unit-cell size in determining the band gap for these organic cation mixtures. Our calculations indicate that the key role played by hydrogen bonds with iodine anions in pure compounds, MAPbI$_{3}$ and FAPbI$_{3}$, is preserved in the cation mixed perovskites. It is found that MA-I bonds remain stronger than FA-I bonds throughout the composition range regardless of the unit-cell expansion as the FA content increases.
                Our calculations reveal how the hydrogen bonds stabilize the no-bonding I-\textit{5p} orbitals, spatially perpendicular to the Pb-I-Pb bond axis, lowering their energy when the H-I interaction occurs, 
                which would explain the well-known role of hydrogen bonding in the structural stabilization of hybrid perovskites. These results contribute to the understanding on the role played by cation mixing at A sites in the physics of lead halide perovskites.} \\

\end{@twocolumnfalse} \vspace{0.6cm}


\renewcommand*\rmdefault{bch}\normalfont\upshape
\rmfamily
\section*{}
\vspace{-1cm}


\footnotetext{\dag~Electronic Supplementary Information (ESI) available: 
Total-energy curves as a function of the cation rotation for MA and FA.
Inorganic framework response to the ordering of cationic orientations. 
Effective ionic radius and tolerance factors of mixed A-site cation perovskites. 
Mixing energy of FA$_x$MA$_{1-x}$PbI$_3$ perovskites.
PBEsol+SOC band structures of mixed halide perovskites near the $R$ point.
DOS and electronic charge density contours for MA$_{0.375}$FA$_{0.625}$PbI$_3$.
Energy distribution of the pDOS curves projected on I-\textit{5p}${\parallel}$ and I-\textit{5p}${\perp}$ orbitals.
See DOI: 00.0000/00000000.}



\section{Introduction}
Since the discovery of hybrid perovskites-based solar cells in 2009,\cite{kojima2009} 
continuous efforts have led to record efficiencies exceeding 22\%\cite{shin2017,kim2020} 
making them promising candidate for future solar power technology. 
In the organic-inorganic trihalide perovskites of general formula ABX$_3$, 
the A-site is occupied by a monovalent organic cation 
(such as, methylammonium CH$_3$NH$_3$ (MA$^{+}$) or formamidinium CH(NH$_2$)$_{2}^{+}$ (FA$^{+}$)), 
the B-site by a divalent metal cation as Pb$^{2+}$,
and the X-site by a halide anion (I$^{-}$, Br$^{-}$, or Cl$^{-}$).
The PbX$_{3}^{-}$ components form inorganic cage framework of corner-sharing octahedra,
eight of which enclose the A-site cation at the center of each cage.
Composition engineering has emerged as an alternative tool to tune 
the structural,\cite{wu2018,charles2017,kubicki2020} 
electrical, and optoelectronic\cite{yang2016,tombe2017,wu2020}
properties of light-harvesting materials.
\begin{figure*}[h]
	\centering
	\includegraphics[height=7cm]{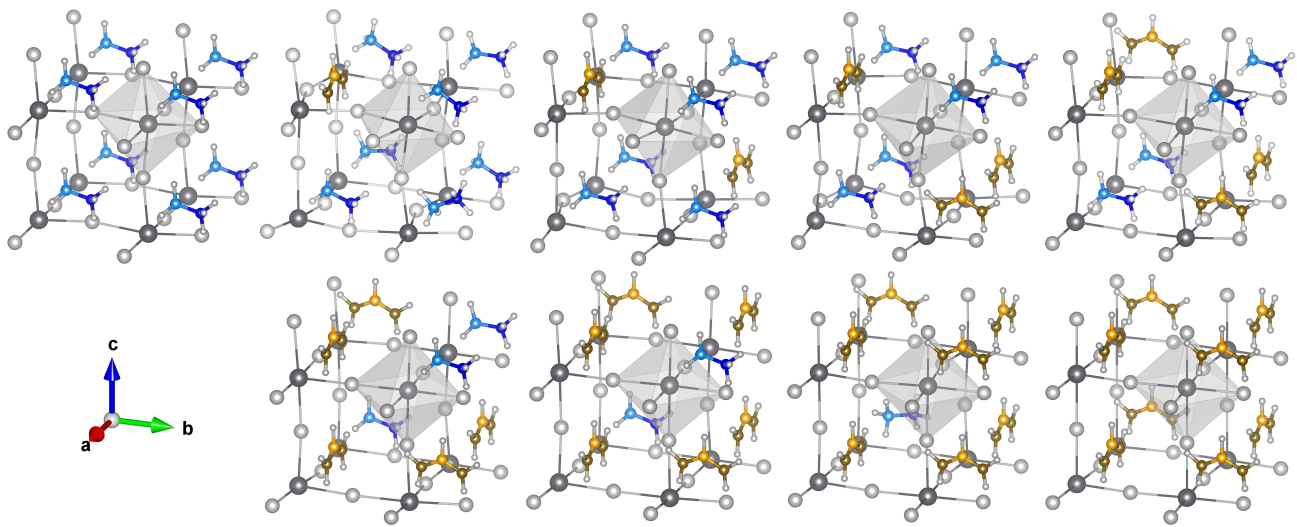}
	\caption{Fully relaxed structures for the FA$_x$MA$_{1-x}$PbI$_3$ solid solution from MAPI (top on the left) to FAPI (bottom on the right), 
		through x= 12.5, 25, 37.5, 50, 62.5, 75, 87.5\%. In organic cations: nitrogen are dark circles and carbon, light circles.}
	\label{fgr:struc}
\end{figure*}
Methylammonium lead triiodide MAPbI$_3$ or MAPI  
has been the most widely investigated hybrid perovskite.
To extend its optical-absorption onset of $\sim$1.61~eV 
further into the red to enhance solar-light harvesting,
Pellet {\it et al.}\cite{pellet2014} proposed a mixture of MA and FA cations 
in the perovskite A-site, which also led to an enhanced short-circuit current 
and thus superior devices to those based only on MA.
In the mixed cation perovskite FA$_x$MA$_{1-x}$PbI$_3$, at room temperature
the band gap varies almost linearly redshifting with increasing the FA content\cite{nazee2017,goni2019}
from the MAPI value until $\sim$1.50~eV in pure FAPbI$_3$ or FAPI. 
In addition, alloying the A-site with organic cations of different size and shape
(the FA molecule is flat, much more symmetrical and larger (effective radius of 2.53~\r{A}) 
than the MA molecule (2.17~\r{A})) also tunes the structural properties.

With decreasing the temperature, MAPI-rich compounds present a phase transition sequence
similar to MAPI: from cubic \textit{Pm$\bar3$m} $\rightarrow$ tetragonal-\textit{I4/mcm} $\rightarrow$ 
orthorhombic \textit{Pnma},
whereas in FA-rich compounds as in FAPI\cite{weberfapi2018} the structural behavior is
cubic \textit{Pm$\bar3$m} $\rightarrow$ tetragonal-\textit{I4/mbm} $\rightarrow$ tetragonal-\textit{P4bm},
as result a rich temperature-composition phase diagram of FA$_x$MA$_{1-x}$PbI$_3$ emerges
with some regions still under discussion.\cite{webermix2016,nazee2017,mohanty2019,goni2020}
From the theoretical perspective, the archetype MAPI and lately FAPI have been extensively 
discussed in the literature.
Properties such as band gap,\cite{leveille2019, bellaiche2017, amat2014}
effective masses,\cite{filip2015, bellaiche2017} 
Rashba splitting,\cite{sukmas2019} hydrogen bonding,\cite{leehbond2016}
optical absorption features,\cite{leveille2019, bellaiche2017} 
phase diagrams,\cite{grozema2019} 
cation interactions,\cite{manzhos2019} 
ionic charges,\cite{madjet2016} 
carrier lifetimes,\cite{filippetti2014} have been modeled.
Mixed cation perovskites instead are less studied and
usually are examined for low concentrations of the substitutional cation.
For instance, ab-initio molecular dynamics calculations 
have shown that in FAPI the substitution of FA$^{+}$  
by a smaller cation like MA$^{+}$ until 10\%
causes an increased tilting of the PbI$_6$ octahedra
to fill in the space left by the smaller cation.\cite{eames2017} 
The same trend has also been observed in FAPI doped with the inorganic cation Cs$^{+}$ 
until 25\%.\cite{islam2018} 
The mixed organic-cation solid solutions FA$_x$MA$_{1-x}$PbI$_3$ are almost theoretically 
unexplored in the complete range of compositions
likely to avoid complications associated with orientations of different type cations.
A better understanding from an atomistic scale about the structure-property relationship 
in these mixed perovskites is necessary in order to contribute improving the performance 
in solar cells that use these materials as a photoactive layer.

The aim of this work is to provide atomic scale information about
the structural and electronic properties of the double organic cation lead halide perovskites.
Based on first-principles calculations, 
we take a static look to the dynamical system FA$_x$MA$_{1-x}$PbI$_3$ with FA content ($x$) 
ranging between the pure end-members MAPI ($x=0$) and FAPI ($x=1$) 
in steps of 12.5\%, assuming cubic cells in all cases.
By inspecting the potential energy surface as a function of the organic molecule orientations
we find a preferential orientation between organic cations.
We characterize then the structural distortions of the Pb/I inorganic sublattice in details,
and address how the interaction organic cation - inorganic framework changes 
in the solid solution with respect to those in the pure-end members. 
Finally, from the analysis of electronic structures
we unravel the key role played by the I-\textit{5p} {\it non-bonding} orbitals 
distributed spatially perpendicular to the Pb-I-Pb bond axis
when saturated with hydrogen bonds.

\section{Computational details}
The FA$_x$MA$_{1-x}$PbI$_3$ solid solution is modeled with a 96-atom supercell 
containing $2\times2\times2$ formula units APbI$_3$ that allows us to consider composition steps $\Delta x$ of 0.125 (12.5~\%). 
%
To tackle modeling of solid solutions with organic cations not symmetrically spherical at the A site, certain approximations must be assumed.
Our first approximation is to set only one particular A-site cation distribution for each composition $x$ as shown in Fig.~\ref{fgr:struc}.
As second approximation, we do not fully examine the cation orientation disorder for each perovskite, instead
the initial relative orientation among organic cations is chosen after inspecting the energy landscape of the pure compounds and of them with a single different substitute cation, as described below and detailed in the ESI\dag.
Third approximation is to keep the simulation supercell cubic for all compositions.
Our first-principles DFT calculations are performed using the generalized gradient approximation (GGA) of Perdew-Burke-Ernzerhof revised for solids (PBEsol)\cite{perdew2008} functional as implemented in VASP\cite{vasp96,vasp99} with an energy cutoff of 550~eV and the Pb-{\textit d}, I, C, N and H projector augmented wave (PAW)\cite{bloch94} potentials provided with the VASP package. The convergence in k-points is obtained with a $\Gamma$-centered $2\times2\times2$ grid. 
%
The atom positions are optimized until the residual forces are smaller than 0.05~eV/\r{A}
maintaining the restrictions described in the next section in each relaxation stage.
All calculations of electronic properties include 
the relativistic spin-orbit coupling (SOC) effect due to the presence of the heavy elements
such as lead and iodine, unless otherwise indicated.

\section{Results and discussion}
\subsection{Structural optimization procedure}

Structural optimization of hybrid metal-halide perovskites is challenging due to 
the soft nature of their potential energy surface.\cite{filip2014} 
The presence of organic molecules which can rotate and displace from the center of the perovskite A-site 
lowers the symmetry and induces tiltings and deformations of the BX$_6$ octahedra.\cite{li2016} 
Furthermore, we find that during structural optimization the orientations of the organic molecules 
hardly change with respect to the initial setting which can be understood by the energy barriers 
they have to overcome between different orientations.
In fact, in cubic MAPI the full energy landscape as a function of the molecular motion 
(rotation plus displacement) scanned through first-principles calculations using one unit cell 
displays energy barriers up to 200~meV.\cite{bechtel2016} 
In FAPI instead the highest energy barriers reported were near 25~meV, 
evaluating Euler's rotations of one FA cation centered in a single unit-cell.\cite{sukmas2019}
All this has led us to design a stepwise optimization procedure to find the most energy-convenient structures.

\begin{figure}[h]
	\centering
	\includegraphics[height=4cm]{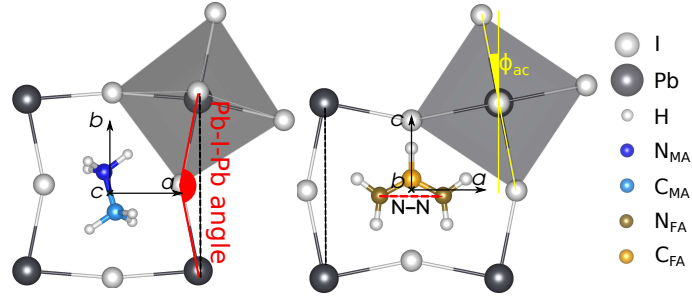}
	\caption{Reference orientation of the organic cations: MA$^{+}$ (left) and FA$^{+}$ (right).
	The tilting angle of octahedra defined as the deviation in Pb-I-Pb angle 
	from the ideal value (180$^{o}$) is indicated in red.}
	\label{fgr:cations}
\end{figure}

For the MA and FA distributions shown in Fig.~\ref{fgr:struc}, 
minimum-energy configurations are obtained following a two stages procedure: 
(1) to determine the most energy-convenient orientation among organic cations,
(2) to perform a complete relaxation of the internal coordinates of all ions.
To begin, the inorganic PbI$_3$-octahedron network is freezed out at
their hypothetical high-symmetry structure whereas only organic cations are moved. 
Cation optimization is in turn divided in two stages: 
(1a) organic cations centered at the A-site of cubic perovskite are rotated
similar to previously reported ``rigid-body'' calculations\cite{bechtel2016,sukmas2019}
to explore the energy surface, and then (1b) they completely relax.

Specifically, our reference structure (space group {\it Pm$\bar3$m}) is defined with
the molecule centered at the A-site (corresponding to the midpoint of
the C-N axis of the MA$^{+}$ cation and the geometrical center of N-C-N
of FA$^{+}$ one) and aligned along the [001] direction (C-N axis and
the N-C-N one, respectively). In stage (1a), inspection of the potential energy surface
when one molecule centered at the A-site rotates around each of the
three Cartesian axis 
is performed for MAPI, FAPI and the mixed compounds with $x$ = 12.5\% and 87.5\%
as shown in Fig.~S1\dag~and S2\dag.
They display that the rotating organic cation prefers 
to orient perpendicularly to the nearest neighbor cations.
In the mixed perovskites analyzed in particular 
the dipole moment of the MA$^{+}$ cation 
points toward the plane defined by a FA$^{+}$ cation.
In fact, this relative orientation between different type cations is used as a guide 
to build the initial configurations for the remaining mixtures.
%
In passing we note that in MAPI, ferroelectric ordering of MA cations 
prevails over the antiferroelectric one in accordance with the dipolar character 
of MA molecules ($\mu$= 2.29~D), 
while in FAPI those orderings of FA cations hardly compete 
in line with FA very small dipole moment ($\mu$= 0.21~D), 
for both compounds our results are consistent with previous similar calculations.\cite{frost2014}
Next, from those lower energy geometries with particular organic cation orientations, 
in stage (1b) a complete molecular relaxation is allowed 
which results in the organic cations moving from their central positions 
in addition to small angle variations in some cases. 
Finally, (stage 2) a complete relaxation of all atoms is performed with the inorganic
network mainly responding to the organic cation arrangements.
Note that all structural optimizations are performed without including SOC, 
which is not expected to have large impact on the geometry of the Pb-based halide 
perovskites.\cite{whalley2017,varadwaj2019}
In addition, geometry relaxations with SOC significantly increases the computational cost
providing no remarkable changes to the local geometry that could affect 
interatomic interactions.

\subsection{Structural properties}
Figure~\ref{fgr:struc} presents the fully relaxed geometries. 
In FAPI (bottom right structure), all FA cations are oriented 
with their C-H bonds along the \textit{c} axis (as seen in Fig.~\ref{fgr:cations}) and
the N-N axes of nearest-neighbor cations perpendicular to each other on the \textit{a-b} plane, 
consequently Pb-I octahedra rotate on that plane $\phi_{ab}\sim 4^{o}$ 
and are in-phase along the \textit{c}-axis. 
This particular structure is distorted with the tetragonal tilt pattern 
($a^{0}a^{0}c^{+}$ in Glazer notation)  
despite we impose a cubic cell for all compositions.
Then the solid solution is formed from FAPI by progressively replacing FA$^{+}$ cations 
with smaller MA$^{+}$ ones oriented along \textit{b} 
trying that they are perpendicular to their FA neighbors and 
also, same type of cations are as far apart as possible in the supercell 
until forming MAPI (top left structure).
In general, large FA$^{+}$ cations hardly move from their previously relaxed positions
forcing the inorganic cages that surround them to expand on the plane where cations lie, 
this is, the dodecahedra elongates along the N-N axis direction of the FA$^{+}$ cations, 
which besides conditions the alignments of its neighbors at the A sites.
When occupied by MA$^{+}$ cations, these accommodate themselves more easily and even rotate a few degrees.

\begin{figure}[h]
	\centering
	\includegraphics[height=6cm]{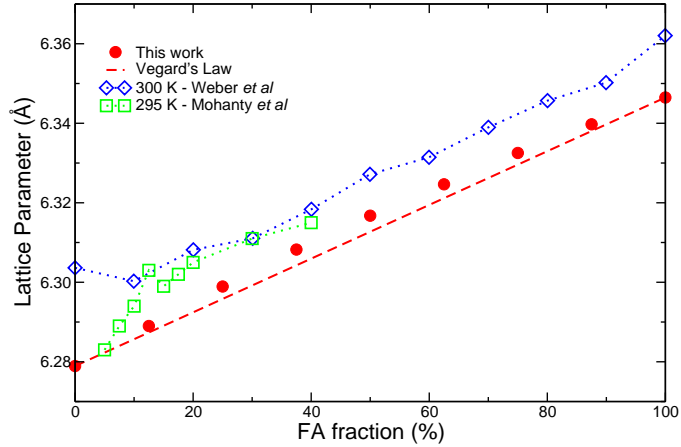}
	\caption{Cubic lattice parameter for the mixed A cation perovskites compared with 
    room-temperature experimental results at 295~K\cite{mohanty2019} and 300~K\cite{webermix2016}.}
	\label{fgr:lattice}
\end{figure}

First, we analyze the structural stability of the mixed perovskites based on the energy balance 
between reactants and products. The mixing energy $\Delta E$ of each solid solution 
with energy $E_j$ is defined as:
\begin{equation}
\Delta E = E_j - x \ E_{FAPI} - (1-x) \ E_{MAPI}
\end{equation}
where the last two terms represent fractions of the total energy of pure compounds. 
Specifically, E$_{FAPI}$ and E$_{MAPI}$ are the total energy per formula unit 
corresponding to the eight-cell minimum-energy configuration. 
The composition dependent mixing energies for the modeled structures (shown in Fig.~S4\dag)
are all negative indicating the improved stability of mixed phases compared to the pure end members. 
As a second criteria to analyze the structural stability, we inspect the Goldschmidt tolerance factor 
which predicts the perovskite phase formation by comparing the size of the A-site organic cation 
with the cavity that contains it. The size-ratio weighted average of the two organic cations MA and FA
is used as the estimated effective cation size $r_{eff}$:
\begin{equation}
r_{eff} = x \ r_{FA} + (1-x) \ r_{MA},
\end{equation}
and the tolerance factor in the mixed FA$_x$MA$_{(1-x)}$PbI$_3$ composition is defined as:
\begin{equation}
t = \frac{r_{eff}+r_{I}}{\sqrt{2}(r_{Pb}+r_{I})}.
\end{equation}
Using the following ionic radii: $r_{MA}$=2.17~\r{A}, $r_{FA}$=2.53~\r{A},
$r_{Pb}$=1.19~\r{A} and $r_{I}$=2.20~\r{A},\cite{webermix2016} 
the tolerance factor for MAPI and FAPI is 0.91 and 0.99, respectively. 
The $t$ values for the FA$_x$MA$_{1-x}$PbI$_3$ solid solutions
present a continuum variation between those pure compound values (Table~1\dag).
In general ABX$_3$ perovskite-like structures have tolerance factors in 
the range $0.8 \le t \le 1.06$,~\cite{franchini2020} so the values for these cation mixtures 
indicate stability of the perovskite structure.

The cubic lattice parameters for the optimized FA$_x$MA$_{1-x}$PbI$_3$ solid solutions 
decrease linearly together with the FA content going from 6.346~\r{A} in FAPI until 6.279~\r{A} in MAPI
as seen in Figure~\ref{fgr:lattice} where the lattice fitting uses the Vegard's law. 
They follow the experimental behavior of the average lattice parameters 
measured at room temperature,\cite{mohanty2019,webermix2016} 
where MAPI and MA-rich perovskites with x<20\% present a tetragonal phase\cite{goni2020}
and the rest a cubic phase.
Our computed zero-temperature values as expected are underestimated.
It would be noted that if the cubic-cell restriction were lifted in the ab-initio calculations, 
each pure end-member would evolve to its low-temperature minimum-energy symmetry, 
namely MAPI to the orthorhombic {\it Pnma} phase and FAPI to the tetragonal phase belonging 
presumably to the {\it P4bm} symmetry group.\cite{goni2020} 
To model then mixed cation solid solutions, for each composition $x$ 
the low-temperature minimum-energy symmetry should first be determined, 
to apply next our optimization procedure, 
which would allow to investigate the composition phase diagram at low temperature.
However, that is out of the scope of the present study and since we aim to explore 
the local variation of the structural and electronic features with the A-site cation mixture, 
a cubic supercell for all mixed systems is a reasonable approximation.
Besides, it is widely used in the literature 
to model hybrid perovskites.\cite{bechtel2016,li2016,zhang2018,zunger2019,zunger2020}

\begin{figure}[h]
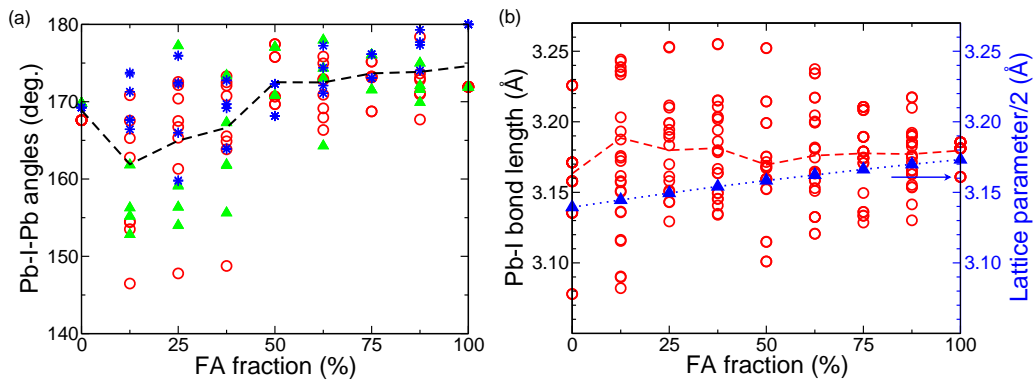

\centering
\includegraphics[height=5cm]{figure_4a}
\includegraphics[height=5cm]{figure_4b}
\caption{ (a) Lead-iodine-lead bond angle for mixed A cation perovskites along 
the three Cartesian directions shown with different symbols. 
The tilting angle of octahedra defined as the deviation in Pb-I-Pb angle 
from the ideal value (180$^{o}$) is marked in Fig.~\ref{fgr:cations}. 
Dashed line joins the average values of each system.
(b) All Pb-I distances (open circles) for each system, whose average values 
are joined with the dashed line. Lattice parameters divided by two are displayed 
with triangles to compare with.}
\label{fgr:angle}
\end{figure}

To analyze the structural properties, we focus first on the alloying effect 
on the lead-halide backbone as it provides the electronic structure of the band edges 
and hence determines the optical properties of hybrid perovskites.
Volume reduction by incorporating smaller MA$^{+}$ cations into FAPI 
directly affects lead-halide octahedra through distortions such as rotations and deformations,
and ultimately by modifying their sizes.
Fig.~\ref{fgr:angle}(a) displays the composition dependence of the octahedron-tilting angles 
defined as the deviation in Pb-I-Pb angle from the ideal angle (180$^{o}$) 
along each Cartesian direction as depicted in Fig.~\ref{fgr:cations}.
They measure the octahedron rotations between nearest neighbors.
From FAPI, the gradual substitution of FA$^{+}$ with MA$^{+}$ cations causes the tiltings on 
average increase smoothly in order to compensate for the reduced space filling 
offered by the smaller MA$^{+}$ cations, 
in agreement with calculations described above.\cite{eames2017}
Instead, MA-rich systems present a pronounced tilting of the corner-sharing octahedra
indicating that larger FA$^{+}$ cations distort much more the smaller MAPI lattice.
In the 50\% system in particular, the A-site disorder expected to be maximum
should increase the octahedron tilting as approaching to that composition 
from both sides in the alloy diagram,
however that does not occur here because of 
the chosen high-symmetry cation distribution.
Figure~\ref{fgr:angle}(b) shows the composition dependence of 
the iodine-lead bond lengths
to analyze the size and deformations inside the octahedra.
In pure FAPI, Pb-I distances are close to the cubic lattice parameters but
with different values,
according to the expansion induced by the in-plane octahedron rotation mentioned above.
As the FA content decreases up to $x$=50\%, 
on average the Pb-I bond lengths (dashed line) 
behave fairly regular around 3.18~\r{A}, above half of the lattice parameters,
because of the slight increase in the rotation of the octahedron
as the cell volume compresses.
From $x$<50\% to MAPI, the lengths of the Pb-I bonds rise above 
the value of 3.18~\r{A} following the behavior of the marked octahedral tilting.
In pure MAPI in particular there is an extra 0.2~\r{A} coming from the used structure 
that has all MA$^{+}$ cations pointing in the same direction
causing the Pb$^{2+}$ cations to displace from the center of the octahedra,
which dilutes as the MA content decreases.

\begin{figure}[h]
	\centering
	\includegraphics[height=7cm]{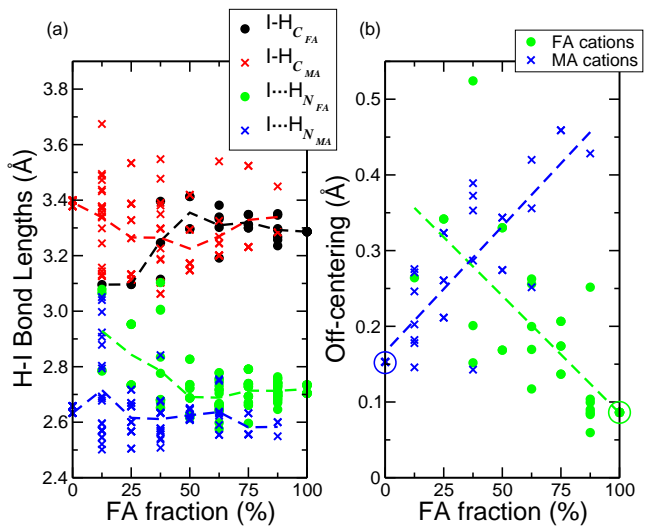}
	\caption{(a) Lengths of iodine-hydrogen bonds and (b) cation off-centering in the optimized structures
		as a function of the alloy composition. Crosses and circles indicate they belong to MA or FA cations, respectively. Different colors in (a) distinguish hydrogen atoms bonded to nitrogen (H$_N$) or carbon atoms (H$_C$).}
	\label{fgr:bond-off}
\end{figure}

We now turn to examine the local interaction between organic cations and the inorganic framework.
In both parent compounds, it is known that they are bonded by hydrogen bonds 
that tether the molecules to the halide anions,
so we measure the hydrogen - iodine distances for each \textit{x} composition 
as shown in Fig.~\ref{fgr:bond-off}a. 
As expected, in all compounds hydrogen atoms bonded to nitrogen atoms H$_N$ are closer 
to iodine anions than those to carbon atoms H$_C$ 
as seen in the two well-separated groups of lengths. 
In the former group,
the H$_N$ belonging to MA cations (H$_{N_{MA}}$) are the closest to the iodine anions 
at an average distance of $\sim$2.6~\r{A} as in pure MAPI
and they notably hold unchanged as the MA ratio decreases in the mixed perovskites 
in spite of the cell expansion and the decrease of the octahedron tilting.
Instead, those belonging to FA cations (H$_{N_{FA}}$) on average form bonds
a bit longer, 2.7~\r{A} in FA-rich compounds, 
but tend to stretch out in the FA-low compounds.
The fact that the I$\cdots$H$_{N_{MA}}$ bonds keep always shorter than 
the I$\cdots$H$_{N_{FA}}$ ones is due to the known feature that MA cations 
are stronger proton donors than FA.\cite{kubicki2020}
In the upper group of I-H$_C$ distances,
the main difference is the scatter of the data which can be understood in terms of the
steric size and shape of the molecular cations.
Unbounded H$_{C_{MA}}$ stay further apart from iodine anions 
and at very different distances, depending on the particular C-N axis orientation;
while H$_{C_{FA}}$ are much closer to unbounded I as 
planar FA cations have hydrogen bonds on both sides and further
occupy a larger area within the surrounding inorganic cages.

The I$\cdots$H$_N$ bond strength results from both 
the displacement of the inorganic sublattice as well as the organic cations.
Fig.~\ref{fgr:bond-off}b displays the cation off-centerings 
measured with respect to the reference structure defined above. 
First, note the vast difference in the pure end-members:
FA cations are barely off-centered ($\sim$ 0.08~\r{A}) in FAPI
compared to the wide displacements of MA ($\sim$ 0.15~\r{A}) in MAPI.
In the solid solutions, we find that the off-centering of each type of cation 
is larger than the corresponding value in its own parent compound,
specifically, starting from the mentioned values they increase 
on average almost linearly as its content in the mixed perovskites decreases.
An extreme case is observed in the mixed perovskite of $x$=87.5\%, 
in which the MA off-centering reaches 0.45~\r{A} being the only one displaced so 
that the length of the I-MA bond remains similar to that in pure MAPI 
(although the alloy volume is 3\% larger), 
since the iodine octahedral only tilt slightly.
Neutron powder diffraction measurements of MAPI\cite{weller2015} indicate that MA cations 
displace towards central positions in the unit cell as the temperature raises.
Our first-principles calculations performed at zero temperature likely overestimate cation off-centerings,
however, we remark that the opposite trend off-centering/alloy fraction
observed in the FA$_x$MA$_{1-x}$PbI$_3$ solid solution  
could be experimentally examined at low temperatures.

Finally, we want to highlight that all optimized systems show that 
the inorganic-framework deformation depends sensitively on the orientation of the organic cations
in agreement with previous results for pure compounds.\cite{li2016} 
To illustrate this point, we compare two completely relaxed configurations 
of pure FAPI (Fig.~S3\dag) 
that were initialized with the FA cations oriented differently:
(i) all parallel and (ii) perpendicular to each other.
After relaxation, the former presents no octahedra rotations at all,
unlike the latter (which is our lower energy FAPI structure)
showing an octahedron tilt pattern $a^{0}a^{0}c^{+}$ as already described.
These structural differences of the inorganic framework can be rationalized 
by examining the organic-inorganic interaction through the hydrogen bond distributions
(as done in Fig.~S3(a)\dag and Fig.~S3(b)\dag).
A similar analysis can be done in the mixed perovskites 
whose hydrogen bonds certainly depend on the type of cation as seen.

\begin{table*}
	\small
	\caption{\ Calculated pseudo-cubic lattice parameters and energy band gaps for FA$_x$MA$_{1-x}$PbI$_3$ 
		using PBEsol and PBEsol$+$SOC. Available experimental data is shown for comparison }
	\label{tbl:gaps}
	\begin{tabular*}{\textwidth}{@{\extracolsep{\fill}}lccccl}
		\hline
		FA fraction & \multicolumn{2}{c}{Lattice parameter (\r{A})} &  \multicolumn{3}{c}{Energy band gap (eV)} \\
		$x$ (\%)    & PBEsol & Expt.                               & PBEsol &  PBEsol$+$SOC & Expt. \\
		\hline
		0    & 6.279 & 6.276\cite{baikie2013}, 6.303\cite{webermix2016} & 1.494  & 0.552  & 1.52\cite{li2017},1.55\cite{leguy2015},1.60\cite{brauer2020} \\
		12.5 & 6.289 & 6.303 \cite{mohanty2019}& 1.611  & 0.641   &  \\
		25   & 6.299 & & 1.528  & 0.505   &  \\
		37.5 & 6.308 & & 1.514  & 0.449   &  \\
		50   & 6.317 & 6.327 \cite{webermix2016} & 1.380  & 0.291   &  \\
		55   &       & &        &         & 1.46\cite{li2017}  \\
		62.5 & 6.325 & & 1.405  & 0.314   &  \\
		75   & 6.333 & & 1.384  & 0.275   &  \\
		87.5 & 6.340 & 6.439\cite{chen2019} & 1.370  & 0.243  & 1.66\cite{chen2019} \\
		100  & 6.346 & 6.362\cite{webermix2016},6.365\cite{jiang2018} & 1.376  & 0.246   & 1.45\cite{li2017}, 1.47\cite{ma2017}, 1.52\cite{prasanna2017} \\
		\hline
	\end{tabular*}
\end{table*}

In summary, the few configurations investigated here indicate
that in mixed cation perovskites, 
while the cell volume depends on the size of the average cation, 
deformation of the inorganic sublattice PbI$_6$ 
is influenced by the orientation of both types of organic cations but 
is more sensitive to the size of the cation with lower content in the system.
The strength of the iodine - hydrogen bonds is less affected by the cation mixture, 
remaining similar to that of each type of cation in its respective parent compound.

\subsection{Electronic properties}
\begin{figure}[h]
	\centering
	\includegraphics[height=5cm]{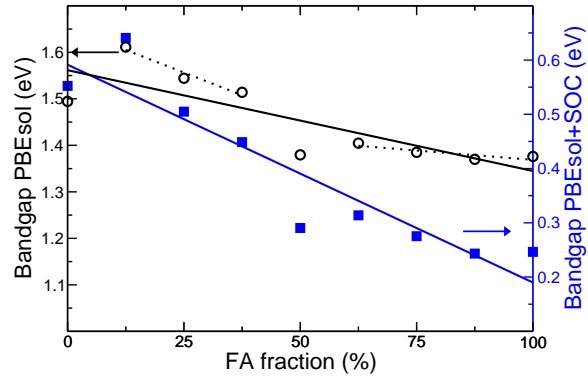}
        \caption{Energy band gaps for mixed A cation perovskites computed using PBEsol (circles) and PBEsol+SOC (squares). 
        Full lines are least-square fits to guide the eyes. 
        Dotted black lines indicate the different slopes of PBEsol band gaps for MA-rich and FA-rich alloys as described in the text.}
	\label{fgr:gap}
\end{figure}

Now we investigate the electronic structure properties of FA$_x$MA$_{1-x}$PbI$_3$.
Band structures of the cation mixed perovskites computed using PBEsol without spin-orbit coupling 
show direct band gaps at the $R$ (1/2, 1/2, 1/2) point of the cubic-phase Brillouin zone.
Band gaps decrease overall linearly as the FA content enhances, 
as shown in Figure~\ref{fgr:gap} (with circles) and listed in Table~\ref{tbl:gaps}. 
They are underestimated with respect to the room-temperature measured optical absorption onsets
of 1.52 - 1.60~eV in MAPI\cite{li2017,leguy2015,brauer2020} and 
1.45 - 1.52~eV in FAPI\cite{li2017,ma2017,prasanna2017} within 0.20~eV.
When the SOC effect is included, calculated energy bands
display a marked splitting in both band edges of pure MAPI (see Fig.~S5\dag),
which dilutes as the FA content rises in the solid solutions
and disappears in pure FAPI.
The Rashba effect observed in this particular configuration of MAPI
has already been reported and assigned to 
the parallel alignment of MA cations that induces a net displacement 
of Pb atoms along the same direction from the octahedron center,\cite{stroppa2015,rubel2018}
breaking in this way the lattice inversion symmetry.
We will return to this point later.
Then, band gaps obtained including SOC present the same decreasing trend
as the FA fraction rises (squares in Fig.~\ref{fgr:gap}) although they are sternly underestimated 
by more than 1~eV compared to experimental values.

\begin{figure*}[h]
	\centering
	\includegraphics[height=8cm]{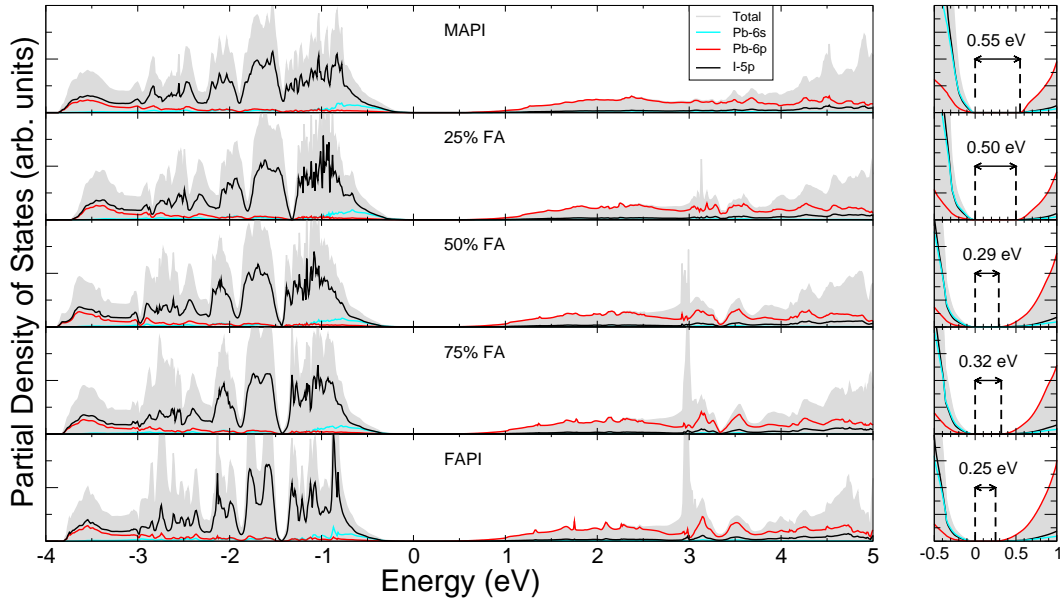}
	\caption{PBEsol+SOC calculated electronic density of states for the FA$_x$MA$_{1-x}$PbI$_3$ 
	systems with the Fermi level sets to zero. Total and projected DOS on I-\textit{5p}, 
	Pb-\textit{6s} and Pb-\textit{6p} (the last two are multiplied by 3). 
	Right: zoom-in around the Fermi level to mark the band gaps.}
	\label{fgr:pdos}
\end{figure*}

\begin{figure*}[h]
	\centering
	\includegraphics[height=7cm]{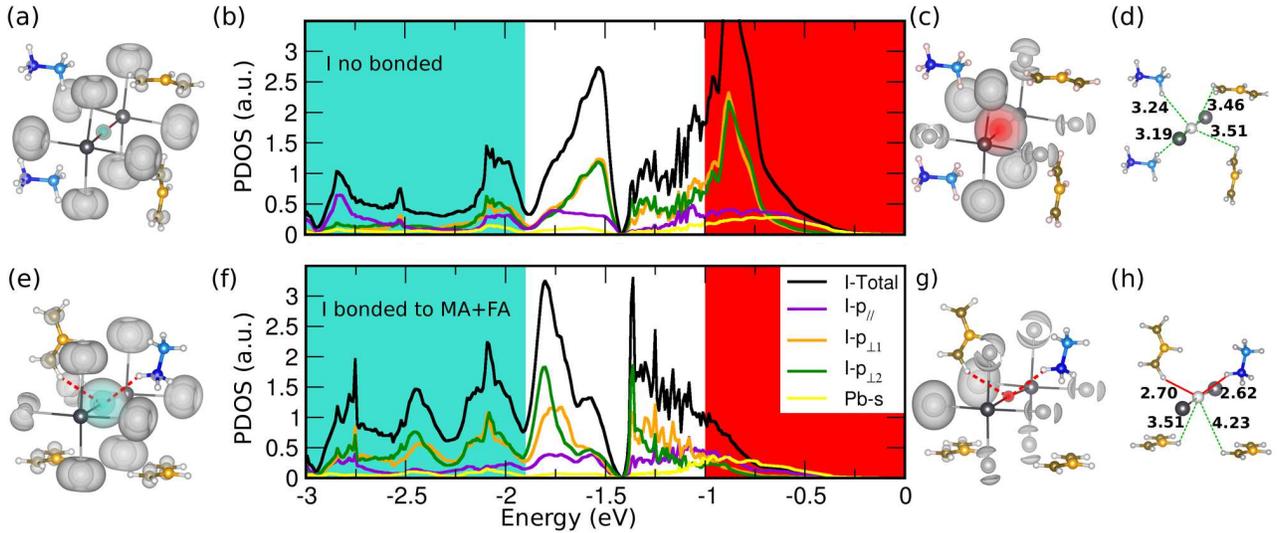}
	\caption{PBEsol+SOC calculated density of states for two selected iodine anions 
in the MA$_{0.375}$FA$_{0.625}$PbI$_3$ solid solution with the Fermi level sets to zero.  
Projected DOS on I-\textit{5p} which is spanned along the parallel (I-\textit{5p}$_{\parallel}$) and 
perpendicular (I-\textit{5p}$_{\perp}$) direction to the Pb-I-Pb bond axis.
(b) For one unbonded iodine and (f) for one iodine bonded to two cations: one MA and one FA.
Electronic charge density contours corresponding to the shaded energy regions of (-3,-1.8)~eV (a,e) and 
(-1,0)~eV (c,g), where only the electronic charge around the selected iodine anions are colored. 
(d,h) Schematic structures of the selected anions with the distances to the closest hydrogen atoms
indicated in Angstrom (\r{A}), and the H-bonding marked in red.}
	\label{fgr:pdos-perp}
\end{figure*}

The electronic properties of lead halide perovskites depend to a great extent 
on the Pb/I framework.
Fig.~\ref{fgr:pdos} displays the total and projected density of states (DOS) 
on Pb-$6s$, Pb-$6p$ and I-$5p$ orbitals for mixed perovskites obtained
including SOC.
Both band edges are determined by antibonding molecular orbitals as in the pure compounds,\cite{lee2016}  
that mainly result from the overlap of I-\textit{5p} and Pb-\textit{6s}  
in the valence band maximum (VBM) and 
from Pb-\textit{6p} and I-\textit{5p} in the conduction band minimum (CBM)
but with predominance of I (Pb) orbitals in the VBM (CBM),
as shown in the figure.
Structural changes in the inorganic sublattice will modify 
the overlap between the Pb-I orbitals and, consequently, the band gap.
An increase in the overlap of the Pb-I orbitals will destabilize the bands enhancing their energies,
and as the VB has greater antibonding character, in general 
the VBM will increase more leading to a net reduction in the band gap.\cite{filip2014,prasanna2017}
The overlap between the Pb-I orbitals is clearly affected by the distance 
and the angle Pb-I-Pb they form.
On one hand, the closer the Pb-I the lower the band gap; 
on the other hand, the octahedron tiltings reduce the overlap between Pb and I
and, consequently, increase the band gap.\cite{filip2014}
Both features 
are sensitive to the cation orientation as we exemplified for two FAPI structures above and further
vary with the MA/FA fraction as we described in Fig.~\ref{fgr:angle},
showing that the effect of octahedron tiltings exceeds 
that of the unit-cell size in determining the band gap of the organic-cation mixture.
This is more noticeable in the band gaps without SOC
that show differences between the MA-rich versus the FA-rich compounds 
(not including the 50\% mixture), as indicated by the dotted lines.
On the other hand, going back to the the Rashba-type splitting observed 
in the band structures, 
from our static DFT calculations we can analyze two contributions: 
the inorganic framework constituted by heavy atoms such as Pb and I, 
and the asymmetrical molecules which ensure inversion symmetry breaking, 
certainly without considering their thermal motion.
As said, conduction band has larger contribution from heavy Pb cations, 
while the valence band is dominated by the lighter iodine anions, 
hence it is expected that the CBM shows a higher Rashba splitting 
than the VBM, such as we have observed in Fig.~S5\dag.
Concerning the organic molecules, 
both the A-site cation mixing and the cation orientational disorder
can break the inversion symmetry. 
Their main final effect is to generate structural deformations in the inorganic PbI$_6$ sublattice.
The relatively strong Rashba effect found in pure MAPI 
which dilutes as the FA fraction increases in the alloys suggests that in our mixed structures,
MA cation ordering plays a major role in the Rashba-type splitting, more than cation mixing.
A previous study on A-site FA/Cs mixtures\cite{islam2018} indicates that
inorganic structural distortions induced by the large cationic size mismatch 
between FA (2.53~\AA) and Cs (1.67~\AA) are the main cause of the band edge splitting
observed in FA$_x$Cs$_{1-x}$PbI$_3$ with $x \le 0.25$.
In the mixed perovskites investigated here, the small difference between cationic radii of MA and FA 
even though leads to inorganic structural deformations (Fig.~\ref{fgr:angle}), 
they are not enough to induce band splitting by themselves.

To further investigate the organic cation effect on 
the electronic structure at the band edges
specifically in the valence band,
we decompose the I-\textit{5p} orbitals 
into three contributions: one projection along the Pb-I-Pb bond axis 
labeled as I-\textit{5p}$_{\parallel}$ and two perpendiculars to that, I-\textit{5p}$_{\perp}$.
To present the results, the solid solution MA$_{0.375}$FA$_{0.625}$PbI$_3$ is selected
as a model case for two reasons: 
(i) the rotations of the octahedra are relatively small ($\sim175^{o}$) 
hence the projections on I-\textit{5p}$_{\parallel}$ and I-\textit{5p}$_{\perp}$ are pretty similar to those 
along the Cartesian axes; and (ii) this system has iodine anions with all kinds of bonds: 
bonded to one and to two cations, of the same type and different, and also \textit{no bonded}. 
The last two cases are illustrated in Fig.~\ref{fgr:pdos-perp}. 
In both cases, we note that I-\textit{5p}$_{\parallel}$ overlaps with the Pb-\textit{6s} orbitals 
in the region close to the BV edge and with the Pb-\textit{6p} (not shown in this figure)
around -3~eV, 
indicative of the covalent character of the Pb-I bonds and in agreement with previous reports 
that analyze the global contribution of the I-\textit{5p} orbitals.
With regards to the I-\textit{5p}$_{\perp}$ orbitals (named \textit{non bonding} 
because they do not actively participate in the Pb-I bond),
we find that they are very sensitive to the interaction with organic cations 
since they are spatially more accessible to the H-N groups.
When one iodine anion is unbound (top figures), its pDOS peak in the energy region 
between -1~eV and the VBM, and 
the corresponding partial charge density presents a spherical charge cloud 
slightly flattened along the Pb-I axis (Fig.~\ref{fgr:pdos-perp}c).
A remarkable difference is observed in bonded iodine anions,
which exhibit a significant energy redistribution of their \textit{non-bonding} orbitals 
increasing in the energy range of -3 to -2~eV at the expense of decreasing in the region near the VBM,
as observed in Fig.~\ref{fgr:pdos-perp}f for a particular iodine bonded to one MA and one FA cation.
Its corresponding electronic charge density (Fig.~\ref{fgr:pdos-perp}e) evidences 
this behavior with a large quasi-spherical cloud over the I 
with their \textit{5p}$_{\perp}$ orbitals saturated,
whereas the charge density of the unbound iodine 
in this energy range is negligible (Fig.~\ref{fgr:pdos-perp}a).
For completeness, iodine cations bounded to one and to two cations of the same specie 
are shown in Fig.~S6\dag.
All these figures reveal how the saturation with hydrogen bonds 
of the \textit{5p} orbitals perpendicular to the Pb-I-Pb bond axis  
stabilizes the \textit{non-bonding} states shifting them to lower energies.
The subtle differences between results for iodine anions bonded to MA against 
those to FA cations do not allow us to conclude about them.
To quantify the redistribution in energy, percentages of the areas under the pDOS curves 
(Fig.~S7\dag) unambiguously show how the I-\textit{5p}$_{\perp}$ are 
totally influenced by the organic cations 
as opposed to the barely affected I-\textit{5p}$_{\parallel}$.
This trend holds for the whole composition range as well as in the pure counterparts
even using another level of theory.

The fact that the inorganic-organic interaction through 
the formation of hydrogen bonds I$\cdots$H$_{N}$ affects 
the electronic state of the iodine atoms has already been observed 
in early works, for example, Mosconi~\textit{et al.}\cite{mosconi2014}
showed that the dynamics of the MA cations in MAPI directly influences the VBM.
Later, Kato~\textit{et al.}\cite{kato2017} noted that the closer I and N are, 
the smaller the valence charge density of iodine atoms 
so they named \textit{anti-coupling effect}.
However, those authors did not project the I-\textit{5p} orbitals and 
their analysis was performed in an energy region near the VBM (between -0.6 and -0.4~eV)
where the I-\textit{5p}$_{\parallel}$ contribution prevails over the I-\textit{5p}$_{\perp}$ one
and further both components decrease when one I anion interacts with organic cations.
Thus they concluded that the anti-coupling effect originates from the reduction of I-\textit{5p} orbitals 
but without unraveling the key role played by the I-\textit{5p}$_{\perp}$ orbitals found here.
Finally, we suggest that the energy lowering 
of the I-\textit{5p}$_{\perp}$ states of bonded halides to organic cations 
can be considered as a feasible mechanism to explain 
the well-known role of hydrogen bonding in the structural stabilization of hybrid perovskites.

\section{Conclusions}
We have employed first-principles calculations to examine the local and 
static properties of the FA$_x$MA$_{1-x}$PbI$_3$ solid solutions as the cation concentration changes.
Assuming a pseudo-cubic cell for all compositions
and modeling only certain structures whose orientation among organic cations 
results of examining the energy landscape, 
the linear variation of the volume and the band gap of the mixed perovskites 
between the two end, MAPbI$_3$ and FAPbI$_3$, is observed 
in agreement with experimental results. 
The cation mixture at the A-site causes local changes in the inorganic PbI$_6$ octahedra
with an increase tilting which varies according to the type of cation with lower concentration. 
In the MA-rich compounds, the replacement of MA by a larger cation FA is to expand 
the space needed for the larger cation, 
whereas in the FA-rich alloys, the replacement of FA by a smaller MA cation is to compensate 
for the reduced space filling offered by the smaller cation.
We confirm the fundamental role played by hydrogen bonds in the mixed perovskites 
as happens in the two parent compounds. 
In agreement with the donor character of the MA cations, 
MA-I bonds remain stronger than FA-I bonds in the whole composition range 
regardless the unit-cell expansion as the FA content increases, 
as result MA cations are significantly off-centered in FA-rich compounds.
The main contribution of this work is to reveal how the hydrogen bonds stabilize 
the no-bonding I-\textit{5p} orbitals, spatially perpendicular to the Pb-I-Pb bond axis, 
lowering them in energy when the H-I interaction occurs.
These results provide insights on the role played by organic cation mixing at A sites 
in the physics of lead halide perovskites.

\section*{Conflicts of interest}
The authors declare no conflict of interest.

\section*{Acknowledgements}
This work is supported by CONICET-Argentina (PUE-0054) and
by Agencia I+D+i Argentina (PICT 2018-01614).
M.S. acknowledges CONICET-Argentina for the Ph.D. fellowship.
This work used computational resources from the Piray\'{u} cluster acquired with funds from ASACTEI, 
Government of the Province of Santa Fe, Argentina (grant AC-00010-18, Res.~117/14). 
The cluster is part of the National System of High Performance Computing of the Ministry of Science 
and Technology of the Rep\'{u}blica Argentina.



\balance


\bibliography{rsc} 
\bibliographystyle{rsc} 

\end{document}